\documentclass[FBSedit,FBSmath,ecsub]{FBSsuppl}
\usepackage{amsfonts}
\usepackage{amssymb}
\include{psfig}

\def \se{$[ 6 ]_{\rm O}$ }
\def \cu{$[ 42 ]_{\rm O}$ }

\title{$NN$ Interaction in Chiral Constituent Quark Models}

\author{A. Valcarce\instnr{1},
F. Fern\'andez\instnr{1},
P. Gonz\'alez\instnr{2}}
\instlist{Grupo de F\' \i sica Nuclear,
Universidad de Salamanca, E-37008 Salamanca, Spain
\and Dpto. de F\' \i sica Te\'orica and IFIC,
Universidad de Valencia-CSIC, E-46100 Burjassot, Valencia, Spain}

\begin{document}

\maketitle
\begin{abstract}
We review the actual state in the description of the $NN$ interaction
by means of chiral constituent quark models. We present a series of relevant
features that are nicely explained within the quark model framework.
\end{abstract}

\section{Introduction}
At the end of the 70's the potentialities of quark model
calculations in low-energy hadron physics were established. In a series
of pioneering works, Isgur and Karl \cite{ISKA} performed a
quite successful exhaustive study
of the baryon spectra and at the same time the short-range repulsion
of the nuclear force was qualitatively explained \cite{NESM}. 
These ideas encouraged several groups to 
undertake the ambitious project of trying to understand the $NN$ interaction
in terms of quark degrees of freedom.

Several authors \cite{TOKI,OKYA,FAFE} found a quantitative explanation
of the $NN$ short-range repulsion based on dynamical effects induced by
the one-gluon exchange (OGE). One would say that there exits no
repulsive potential in the naive sense. The color-magnetic OGE interaction
likes to have as many symmetric pairs in the color-spin space as 
possible. Due to the connection of the spin-color space with the
orbital space as a consequence of
the Pauli principle the orbital
symmetry $[42]_{\rm{O}}$ is preferred. 
But this symmetry has to have a zero in the S-wave.
This zero yields for the phase shift 
the same results as a hard or a soft core.

Based on this success simple quark models of the nuclear force
were constructed. The so-called {\it hybrid quark models} \cite{OKYA}
contained three main ingredients:
(i) A short range part given by the
quark-exchange interaction, (ii)  a long range part
by the one-pion exchange (OPE) potential, and (iii) a medium range
part by a phenomenological potential or a two-pion 
exchange potential. These meson exchange potentials
were considered to simulate the meson cloud surrounding the 
quark core. These models allowed to reproduce the $NN$ scattering and bound
state data. The next conceptual step in order
to understand the $NN$ interaction only
based on quark degrees of freedom 
was the consideration of $(q\bar{q})$
and $(q\bar{q})^2$ excitations generated by off-shell
terms of the Fermi-Breit quark-gluon interaction \cite{FUHE}.
Although the obtained potential acquired an
attractive part in the 0.8$-$1.5 fm range, it was
too weak to bind the deuteron or to fit the extreme
low-energy S-wave scattering.

In the middle 80's quark antisymmetry effects
on the OPE started to be analyzed \cite{SHIM}.
A similar $NN$ short-range repulsion 
to the one provided by the OGE was obtained. 
A first quark-model calculation based on gluon and
pion exchange at the level of quarks was performed \cite{BRFA}.
As the OPE contributes to the $\Delta -N$ mass difference, the rather
big quark-gluon coupling constant used in the previous models 
($\alpha_s > 1$) got nicely reduced to values around 0.5. 
However, it was not obtained enough intermediate
range attraction. To avoid this problem, 
a phenomenological scalar potential at baryonic level
was introduced, returning on this way to a hybrid model.
Besides, its coupling constant
was fitted to a different value for $NN$ $S$-waves, than for
higher angular momentum 
partial waves or for the deuteron \cite{BRFA}.

\section{The Chiral Constituent Quark Model}

In the early 90's the constituent
quark mass was related to the breaking of chiral
symmetry \cite{OBKU,FEVA}. The underlying idea 
is that the constituent quark mass is a consequence
of the spontaneous chiral symmetry breaking.
Then, between the chiral symmetry 
breaking scale ($\Lambda_{\rm CSB}\sim$ 1 GeV)
and the confinement scale,
($\Lambda_{\rm C}\sim$ 0.2 GeV) QCD may be simulated
in terms of an effective theory of constituent quarks
interacting through Goldstone modes. Specifically,
in a Nambu-Goldstone realization of chiral symmetry
within the linear sigma model, there appear two Goldstone boson
fields: the pion and the sigma. The chiral constituent
quark model incorporates 
OPE and one-sigma exchange (OSE) interactions 
of the form,
\begin{eqnarray}
V_{\rm OPE} ({\vec r}_{ij}) & = & 
{\frac{1 }{3}} \, \alpha_{ch} {\frac{\Lambda^2
}{\Lambda^2 - m_\pi^2}} \, m_\pi \, \Biggr\{ \left[ \, Y (m_\pi \, r_{ij}) -
{\frac{ \Lambda^3 }{m_{\pi}^3}} \, Y (\Lambda \, r_{ij}) \right] {\vec \sigma%
}_i \cdot {\vec \sigma}_j +  \nonumber \\
& & \left[ H( m_\pi \, r_{ij}) - {\frac{ \Lambda^3 }{m_\pi^3}} \, H( \Lambda
\, r_{ij}) \right] S_{ij} \Biggr\} \, {\vec \tau}_i \cdot {\vec \tau}_j \, ,
\label{OPE}
\end{eqnarray}
\begin{equation}
V_{\rm OSE}({\vec{r}}_{ij})=-\alpha _{ch}\,
{\frac{4\,m_{q}^{2}}{m_{\pi }^{2}}}{%
\frac{\Lambda ^{2}}{\Lambda ^{2}-m_{\sigma }^{2}}}\,m_{\sigma }\,\left[
Y(m_{\sigma }\,r_{ij})-{\frac{\Lambda }{{m_{\sigma }}}}\,Y(\Lambda \,r_{ij})%
\right] \,,
\label{OSE}
\end{equation}
\noindent $\Lambda$ being  a cutoff parameter and $Y(x)$ and $H(x)$
the standard Yukawa functions. Perturbative features of QCD are
incorporated through the OGE potential, 
\begin{equation}
V_{\rm OGE}({\vec{r}}_{ij})={\frac{1}{4}}\,\alpha _{s}\,
{\vec{\lambda}}_{i}\cdot
{\vec{\lambda}}_{j}\Biggl \lbrace{\frac{1}{r_{ij}}}-{\frac{\pi }{m_{q}^{2}}}%
\,\biggl [1+{\frac{2}{3}\vec{\sigma}}_{i}\cdot {\vec{\sigma}}_{j}\biggr ]%
\,\delta ({\vec{r}}_{ij})-{\frac{3}{{4m_{q}^{2}\,r_{ij}^{3}}}}\,S_{ij}\Biggr
\rbrace\,,
\end{equation}
\noindent where the $\lambda ^{\prime }s$ stand for the color SU(3)
matrices. Finally, a phenomenological confining potential is introduced.
Its detailed radial structure, being fundamental to
study the hadron spectra, is expected to play a minor role for the two-baryon
interaction. To be consistent with meson spectroscopy
it can be taken to be linear
\begin{equation}
V_{\rm CON}(\vec{r}_{ij})=-a_{c}\,\vec{\lambda _{i}}\cdot \vec{\lambda _{j}}%
\,r_{ij}\, ,
\end{equation}
though for simplicity harmonic oscillator quark wave functions
will be used for the $NN$ interaction.

This model is able to reproduce quite nicely the $NN$ scattering
and bound state data \cite{FEVA,VABU}.
Typical values for the parameters are given in Table \ref{table1}.
Other quark-model approaches to the $NN$ interaction are found in the
literature \cite{WAGO}.
\begin{table}[hbt]
\caption{Quark-model parameters.}
\label{table1}
\begin{tabular}{|c|c|c|c|c|c|c|c|c|}
\hline
$m_q$ & $b$ & $\alpha_s$ & $a_c$ & $\alpha_{ch}$ & $r_0$ 
& $m_{\sigma}$  & $m_{\pi}$ & $\Lambda$  \\
MeV & fm & & MeV fm$^{-1}$ & & fm & fm$^{-1}$ & fm$^{-1}$ &
fm$^{-1}$  \\ 
\hline
313 & 0.518 & 0.485 & 67.0 & 0.0269  & 0.25  & 3.42 & 0.7& 4.2 \\
\hline
\end{tabular}
\end{table}
\section{Several Reasons to Use Quark Models}

\subsection{Medium Range Attraction}
\begin{figure}[b]
\mbox{\psfig{figure=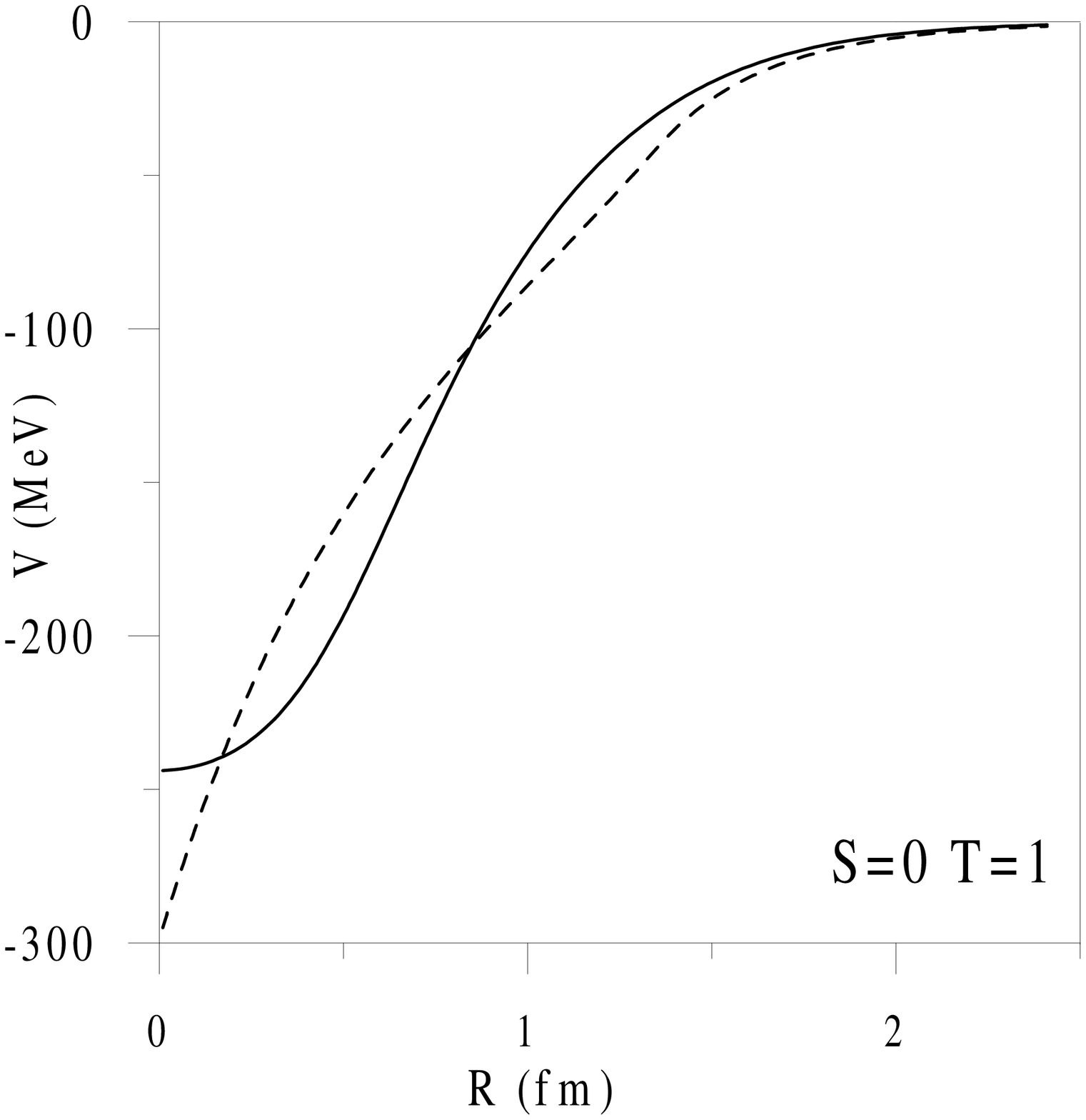,height=2.8in,width=2.2in}}
\hspace*{0.6cm}
\mbox{\psfig{figure=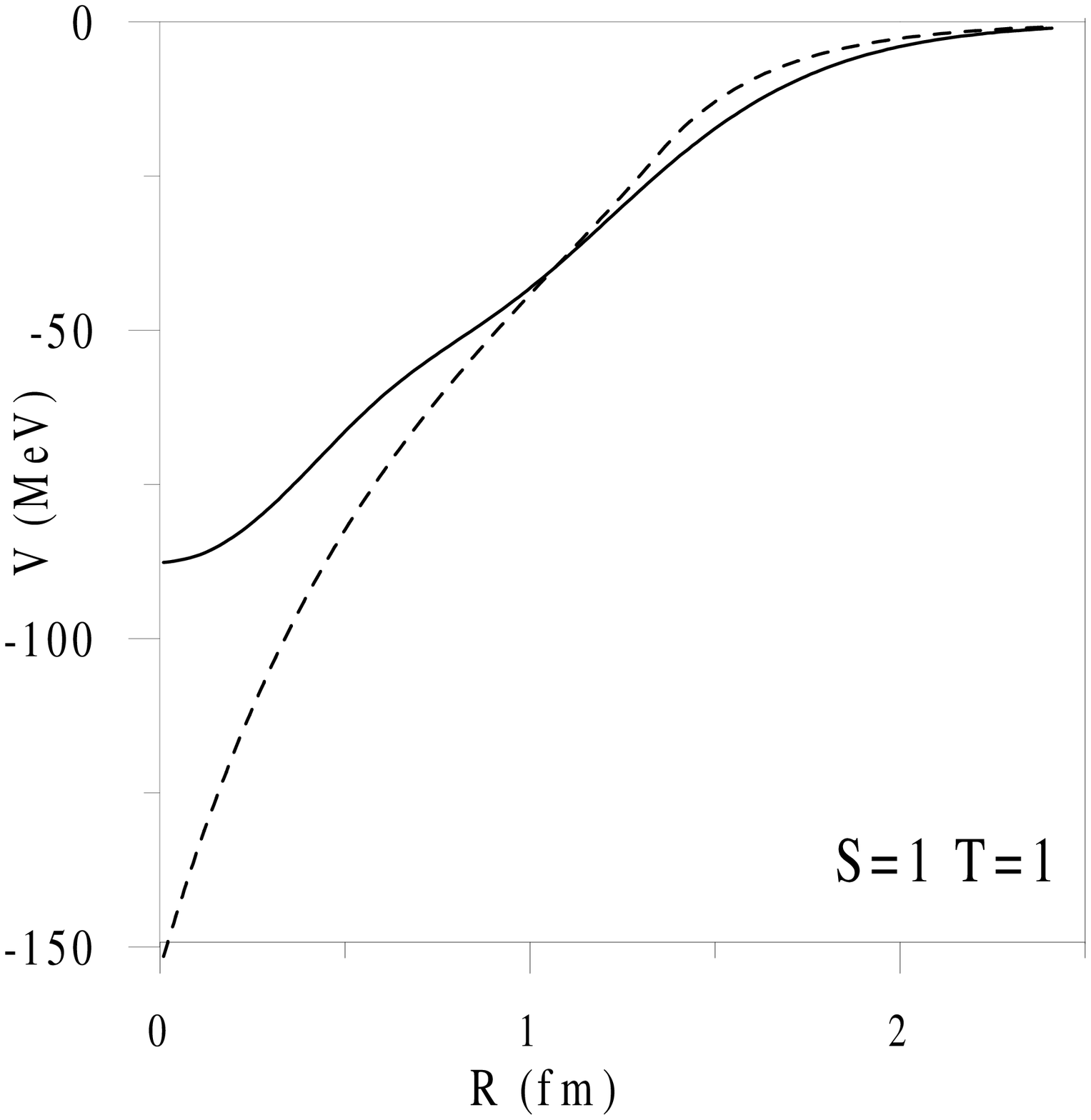,height=2.8in,width=2.2in}}
\vspace*{-3.1cm}
\caption{$NN$ $\sigma$ exchange potential} 
\label{fig1a}
\end{figure}
\begin{figure}[t]
\vspace*{-0.3cm}
\mbox{\psfig{figure=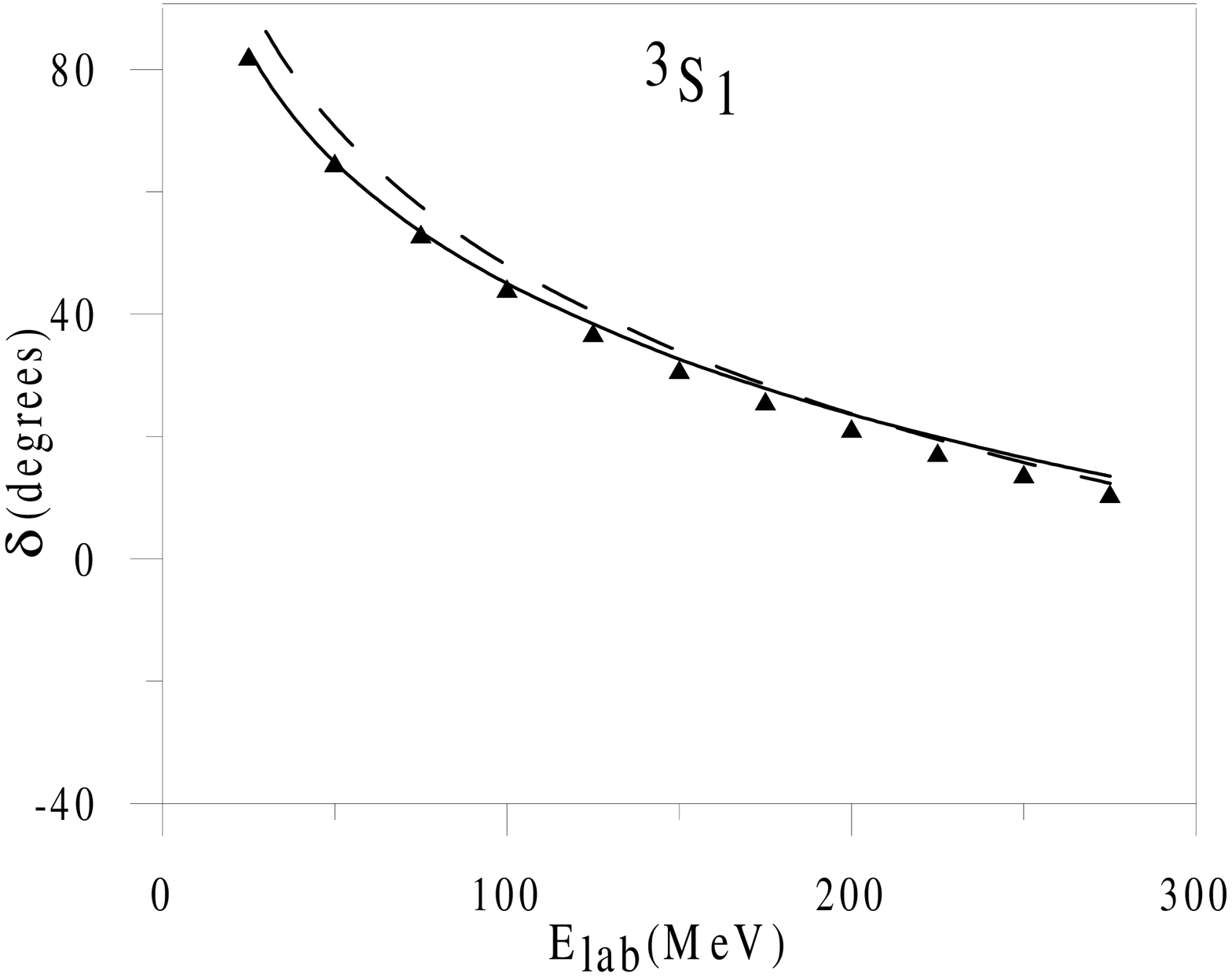,height=3.2in,width=2.2in}}
\hspace*{0.5cm}
\mbox{\psfig{figure=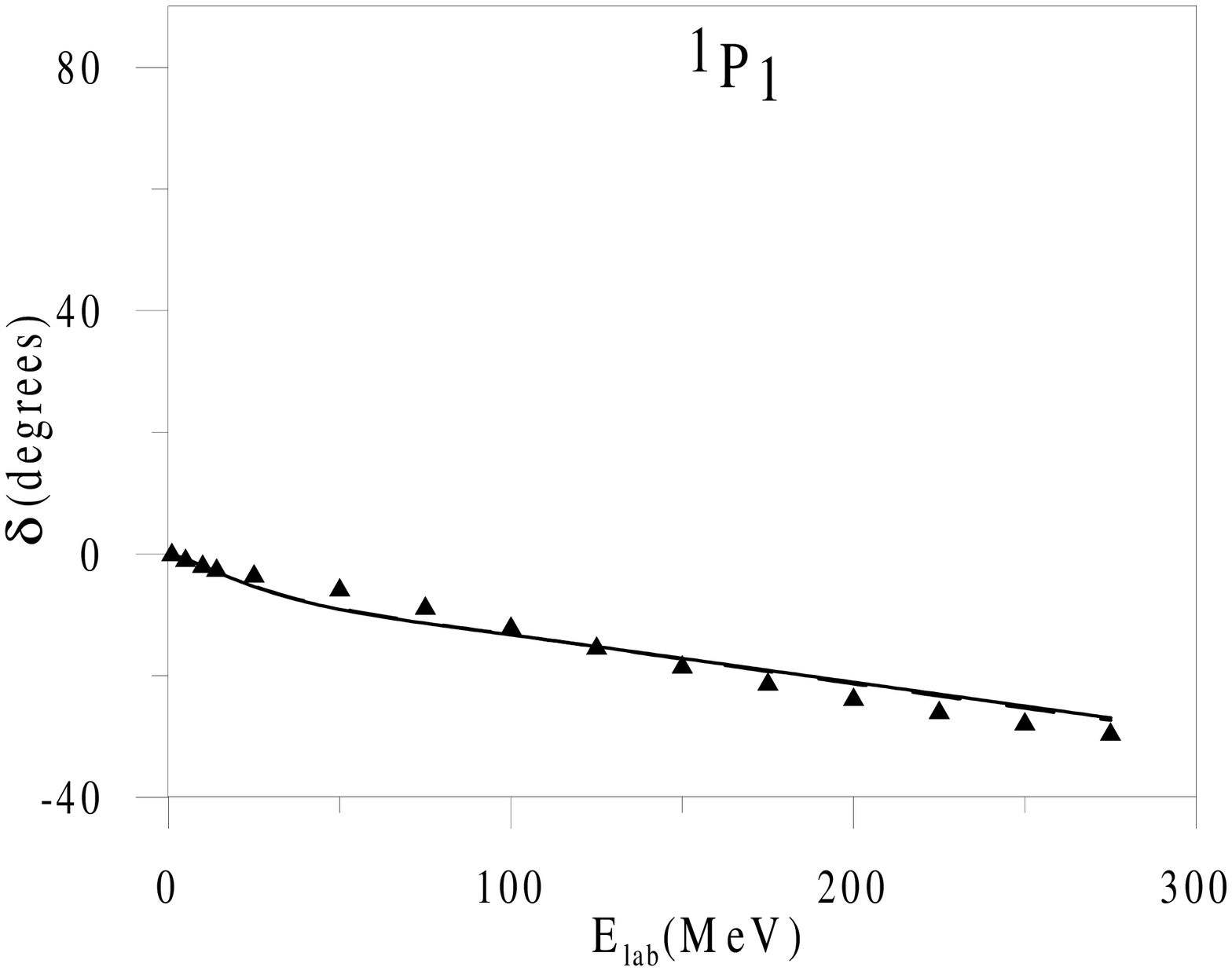,height=3.2in,width=2.2in}}
\vspace*{-3.5cm}
\caption{$^3S_1$ and $^1P_1$ $NN$ phase shifts} 
\label{fig1b}
\end{figure}

A new feature of the chiral constituent quark model is that 
through the OSE potential 
incorporates in a natural way the $NN$ medium range 
attraction without any additional parameter. This fact is
illustrated in Fig. \ref{fig1a} where we compare
the scalar potential obtained at quark level (solid line)
with the baryonic 
parametrization used in ref. \cite{BRFA} (dashed line)
to fit the experimental data.
While at the baryonic level two different 
coupling constants are used:
$g^2_{\sigma NN}/4\pi=$ 3.7 for (S,T)=(0,1) and
$g^2_{\sigma NN}/4\pi=$ 1.9 for (S,T)=(1,1),
the quark model result is obtained in both cases with the
same parameters fixed from the $\pi qq$ coupling constant.
In Fig. \ref{fig1b} we compare the phase shifts for both cases
and we can see how the predictions are similar. 
\subsection{$NN$ Short-Range Repulsion}

The $NN$ short-range behavior has been explained
in terms of an interplay between quark
antisymmetry and a particular quark dynamics.
In a group theory language, a completely antisymmetric
six quark state asymptotically
describing two free nucleons in a relative S-wave is given,
\begin{equation}
\Psi_{\rm NN} = \left[ \, \Psi_{\{ 6 \}} +
\sqrt {8} \, \Psi_{ \{ 42 \} }\right] /3 \, ,
\end{equation}
\noindent
where
\begin{eqnarray}
\Psi_{\{ 6 \}} \, & = & \, [ 2^3 ]_{\rm C}
\,\, [ 6 ]_{\rm O} \,\, [ 33 ]_{\rm ST}
\, , \nonumber \\
\Psi_{\{ 42 \}} \, & = & \, {1 \over \sqrt{2}}
\,\, [ 2^3 ]_{\rm C} \,\,
[ 42 ]_{\rm O} \, \left( [ 33 ]_{\rm ST} \, -
\, [ 51 ]_{\rm ST} \right) \, ,
\label{l1}
\end{eqnarray}

\noindent
the subindex ${\rm C}$, ${\rm O}$ or ${\rm ST}$
indicating the color, orbital
or spin-isospin representation respectively. 
When the distance between the two clusters goes to zero
the $\{ 42 \} $ configuration increases in energy.
According to our harmonic oscillator wave function
ansatz for the quarks, the excitation energy
is given by, 
\begin{equation}
\Delta_{\rm ho} \, \equiv \,
\Delta \left[ E_{\{ 42 \} } - E_{ \{ 6 \} } \right]_{\rm ho}
\, = \,
2 \, \hbar \, \omega \, = \, {{2 \, \hbar^2} \over
{3 \, b^2 \, m_q}} \, .
\label{f2}
\end{equation}
Let us now consider other interactions.
Assume first that quarks interact via the OGE 
used in ref. \cite{OKYA} ($\alpha_s = 1.39$,
$m_q = 300 \,\,{\rm MeV}$ and $b=0.6 \,\,{\rm fm}$).
We can estimate the energy contribution of the OGE
calculating the matrix element,
$< \{ \, \} \mid \sum_{i < j}
V_{\rm OGE} ({\vec r}_{ij}) \, \mid \{ \, \} >$ where
$\{ \, \}$ stands for the
$\Psi_{\{ 42 \}}$ and $\Psi_{\{ 6 \}}$ configurations of
Eq. (\ref{l1}).
As we are interested in the $NN$ interaction energy we
should subtract twice the corresponding nucleon
self-energy given by
$< {\rm N} \mid \sum_{i < j}
V_{\rm OGE} ({\vec r}_{ij}) \, \mid {\rm N}>$. 
For (S,T)=(1,0), the matrix element of the color magnetic potential
for the $\{ 6 \}$ configuration is
48.83  MeV, 
while for the $\{ 42 \}$ configuration
is $-$ 235.63 MeV.
Since in the energy difference between the $\{ 42 \}$
and $\{ 6 \}$ configurations the subtracted self-energies
cancel, we have
$\Delta_{\rm OGE} \equiv
\Delta \left[ E_{\{ 42 \} } - E_{ \{ 6 \} } \right]_{\rm OGE}
\, = \, - 284.46 \,\, {\rm MeV}$.
This difference compensates the
harmonic oscillator energy
difference, $\Delta_{\rm ho} = 240.36 \,\,{\rm MeV}$, the
\cu orbital symmetry becoming the lowest in energy. As
a consequence of this configuration mixing
the S-wave $NN$ relative motion wave function shows
an energy independent node, which translates into
a hard-core behavior in the phase shift.

In the case of the chiral constituent quark model
the value of $\alpha_s$, which drives
the OGE energy gap between the $\{ 42 \}$ and $\{ 6 \}$
configurations, is significantly reduced due to the
pseudoscalar contribution to the $\Delta -N$ mass difference
and correspondingly the mixing effect.
If we recalculate the contribution
of the OGE with the parameters of Table \ref{table1}, we observe again
a decrease of the energy difference between the
$\{ 42 \}$ and the $\{ 6 \}$ configurations
$\Delta_{{\rm OGE}} = \, - 140.99 \,\, {\rm MeV}$,
but it is much smaller
than the harmonic oscillator energy difference
$\Delta_{\rm ho} = \, 309.10 \,\, {\rm MeV}$
and it is not sufficient to give rise to a significant
configuration mixing.
It is then convenient to deep in the
understanding of the repulsion mechanism.

A similar kind of analysis can be carried out for the OPE
and OSE potentials.
For (S,T)=(1,0), differently than in the
OGE case, the OPE has the same sign for both configurations.
Regarding the energy difference one gets
$\Delta_{\rm OPE} = - 68.62 \,\, {\rm MeV}$.
Concerning the OSE one has
$\Delta_{\rm OSE} = 133.05 \,\, {\rm MeV}$.
Therefore, in the chiral constituent quark model three effects
conspire against the symmetry mixing. First, the small value
of $\alpha_s$. Second, the cancellation between the
OPE contributions to the $\{ 42 \}$ and $\{ 6 \}$
configurations, and finally the contrary effect of the
OSE potential. Putting all the contributions together one obtains,
$\Delta \left[ E_{ \{ 42 \} } - E_{ \{ 6 \} }
\right]_{{\rm OGE} + {\rm OPE} + {\rm OSE}} \,
= - 76.56 \,\, {\rm MeV}$,
which is much smaller than the
harmonic oscillator energy difference
$\Delta_{\rm ho} = 309.10 \,\, {\rm MeV}$.
Then, one must conclude 
that in the chiral constituent quark model
there is not enough configuration mixing to account for the
hard-core of the $NN$ interaction as a node produced by the
\cu orbital symmetry.

To look for the origin of
this repulsive character of the interaction one should go beyond
the energy difference between the symmetries and calculate the
specific contribution of the interaction for each symmetry.
In order to see the repulsive or
attractive character of every term of the potential in both
configurations one has to subtract twice the
corresponding nucleon self-energy, given by
${\cal E}_{\rm OGE} = -72.63 \,\, {\rm MeV}$,
${\cal E}_{\rm OPE} = -311.40 \,\, {\rm MeV}$, and
${\cal E}_{\rm OSE} = -66.90 \,\, {\rm MeV}$.
One then obtains,
${\rm E}_{ \{ 6 \} }^{{\rm OGE} + {\rm OPE} + {\rm OSE}} \, = \,
455.72 \,\, {\rm MeV}$
and 
${\rm E}_{ \{ 42 \} }^{{\rm OGE} + {\rm OPE} + {\rm OSE}} \, = \,
379.16 \,\, {\rm MeV}$.
It is then clear that the chiral potential produces strong
repulsion in both configurations, $\{ 42 \}$ and
$\{ 6 \}$. The OPE hardly contributes to the mixing between them,
however in both symmetries it produces strong repulsion.
The OGE reduces the energy difference
whereas the OSE increases this energy difference and produces
an additional attraction in both configurations. The whole
effect is an energy difference of about the same value obtained
with the harmonic oscillator model but with an additional repulsion in both
symmetries originated mainly by quark antisymmetry on the OPE.
Therefore, in this type of models the $NN$ S-wave behavior
should be attributed to the strong repulsion generated by the OPE in the
\se orbital configuration.

\subsection{Universality}

A main feature of the quark treatment is its universality in the sense that
all the baryon-baryon interactions are treated on an equal footing.
Moreover, once the model parameters are fixed from $NN$ data
there are no free parameters for any other case. This allows a
microscopic understanding and connection of the different baryon-baryon
interactions that is beyond the scope of any analysis based only on
effective hadronic degrees of freedom. 
We will illustrate our discussion by means of the recently
calculated $NN \to NN^*(1440)$ transition potential \cite{JUGO}. 
In particular we will determine
the $\pi NN^{\ast }(1440)$ and $\sigma NN^{\ast }(1440)$ coupling
constants.

For this purpose, let us realize that asymptotically
($R \geq 4$ fm) the OSE and OPE potentials have at the baryon
level the same spin-isospin structure than at the quark level.
Hence we can parametrize the asymptotic central interactions as, 
\begin{eqnarray}
V_{NN\rightarrow NN^{\ast}(1440)}^{OPE}(R) & = &\frac{1}{3} \, 
\frac{g_{\pi NN}}{%
\sqrt{4\pi }} \, \frac{g_{\pi NN^{\ast}(1440)}}{\sqrt{4\pi }} \, \frac{%
m_{\pi }}{2M_{N}} \, \frac{m_{\pi }}{2(2M_{r})} \, \frac{\Lambda ^{2}}{%
\Lambda ^{2}-m_{\pi }^{2}}  \nonumber \\
& & [(\vec{\sigma }_{N}.\vec{\sigma }_{N})(\vec{\tau }%
_{N}.\vec{\tau }_{N})] \, \frac{e^{-m_{\pi }R}}{R} \, ,  \label{lrg}
\end{eqnarray}
\noindent and
\begin{equation}
V_{NN\rightarrow NN^{\ast}(1440)}^{OSE} (R)=- \, \frac{g_{\sigma NN}}{\sqrt{%
4\pi }} \, \frac{g_{\sigma NN^{\ast}(1440)}}{\sqrt{4\pi }} \, \frac{\Lambda
^{2}}{\Lambda ^{2}-m_{\sigma }^{2}} \, \frac{e^{-m_{\sigma }R}}{R} \, ,
\label{slrg}
\end{equation}
\noindent where $g_{i}$ stands for the coupling constants at the baryon
level and $M_{r}$ is the reduced mass of the $NN^{\ast }(1440)$ system. 

By comparing these baryonic potentials with the asymptotic behavior of the
OPE and OSE obtained from the quark calculation we can extract
the $\pi NN^{\ast }(1440)$ and $\sigma NN^{\ast }(1440)$ coupling constants
in terms of the elementary $\pi qq$ coupling constant 
and the one-baryon model dependent
structure. The sign obtained for the meson-$NN^{\ast }(1440)$ coupling
constants comes determined by the arbitrarily chosen
relative sign between the $N$ and $N^{\ast }(1440)$ wave functions. Only
the ratios between the $\pi NN^{\ast }(1440)$ and $\sigma NN^{\ast }(1440)$
would be free of this uncertainty. 

To get $g_{\pi NN^*(1440)}/\sqrt{4\pi }$ we turn to our numerical
results for the $^{1}S_{0}$ OPE potential, and fit
its asymptotic behavior to Eq. (\ref
{lrg}). We obtain
\begin{equation}
\frac{g_{\pi NN}}{\sqrt{4\pi }} \frac{g_{\pi NN^{\ast}(1440)}}{\sqrt{4\pi }}
\frac{\Lambda ^{2}}{\Lambda ^{2}-m_{\pi }^{2}}= \, - \, 3.73 \, ,
\end{equation}
\noindent i.e. $g_{\pi NN^{\ast }(1440)}/\sqrt{4\pi}= - 0.94$. As
explained above only the absolute value of this coupling constant is well
defined.
The coupling scheme dependence can be explicitly eliminated if we compare 
$g_{\pi NN^{\ast }(1440)}$ with $g_{\pi NN}$ extracted from the 
$NN\rightarrow NN$ potential within the same quark model approximation.
We get
\begin{equation}
\left | g_{\pi NN^{\ast}(1440)}/g_{\pi NN} \right |=0.25 \,.
\label{eq17}
\end{equation}
By proceeding in the same way for the scalar potential
we can write
\begin{equation}
\left |g_{\sigma NN^{\ast}(1440)}/g_{\sigma NN} \right |=0.47 \, .
\label{eq18}
\end{equation}

The ratio given in Eq. (\ref{eq17}) is similar to that obtained in ref. \cite
{RIBO}. Nonetheless one can find
values for $f_{\pi NN^{\ast }(1440)}$ ranging between 0.27$-$0.47
coming from different experimental analyses.
Regarding the ratio obtained in Eq. (\ref{eq18}), our result agrees quite
well with the only experimental available result, obtained in ref. \cite
{HIOS} from the fit of the cross section of the isoscalar Roper excitation
in $p(\alpha,\alpha^{\prime})$ in the 10$-$15 GeV region, where
a value of 0.48 is given.
Finally, we can give a very definitive prediction of the magnitude and
sign of the ratio of the two ratios,
\begin{equation}
g_{\pi NN^{\ast}(1440)}/g_{\pi NN}=0.53 \; g_{\sigma
NN^{\ast}(1440)}/g_{\sigma NN} \, .
\end{equation}

\subsection{Pauli Blocking}

The Pauli principle plays an essential role to explain the atomic
and nuclear structure. For example,
the strong repulsion existing in the $\alpha \alpha$ system 
originates from the nucleonic substructure of the $\alpha$
particle and not from a 
particular dynamics. The hard-core appears
because the saturation of nuclear spin-isospin
degrees of freedom allows only four nucleons in the
orbital ground state. We shall refer
to this saturation phenomenon indicated by the
existence of forbidden states at the underlying
structure level as {\it Pauli blocking}.
At the level of subnuclear physics
the elementary constituents, the quarks, are fermions so
a quark Pauli principle is active. 

For identical baryons the symmetrization postulate
at the baryonic level gives rise to selection
rules on the state of the compound system.
At the quark level,
assuming SU(2) flavor symmetry, this selection
rule appears as a particular case of 
the action of the quark antisymmetrizer,
${\cal A}$, on a two cluster state
from six identical quarks.
Taking into account that any two-baryon
state (LST) can be decomposed in a symmetric
plus an antisymmetric part under the exchange
of the baryon quantum numbers, one can write
for a definite symmetry (specified by $f$)
\begin{eqnarray}
\Psi_{B_1 B_2}^{{\rm LST}}({\vec R}) & = & {{\cal A} \over {\sqrt{1 +
\delta_{B_1 B_2}}}} \sqrt{1 \over 2} \Biggr\{ \left[
\Phi_{B_1} \left( 123 ;{-{{\vec R} \over 2}} \right)
\Phi_{B_2} \left( 456 ; {{\vec R} \over 2} \right)
\right]_{{\rm LST}} \, + \nonumber \\
& + &(-1)^{f} \,
\left[
\Phi_{B_2} \left( 123 ; {-{{\vec R} \over 2}} \right)
\Phi_{B_1} \left( 456 ; {{\vec R} \over 2} \right)
\, \right]_{{\rm LST}} \Biggr\} \,
,\label{Gor}
\end{eqnarray}

\noindent
where $B_1$ and $B_2$ denote the baryons, and $-{{\vec R} \over 2}$,
${\vec R} \over 2$ are the positions of the clusters.
The antisymmetrizer can be decomposed as,
\begin{equation}
{\cal A} \, = {1 \over 2} \, (1-9P_{36})(1-{\cal P}) \, ,
\end{equation}
\noindent
where $P_{36}$ exchanges quarks 3 and 6, and ${\cal P}$
exchanges the two clusters.
The $1-{\cal P}$ factor implies
${\rm L+S_1+S_2-S+T_1+T_2-T+}f \, = \, {\rm odd}$.
For non-identical baryons this relation indicates the
symmetry associated to a given set of values of (LST).
For identical baryons, $B_1=B_2$, $f= {\rm even}$ (in order
to have a no vanishing wave function) and
${\rm L+S+T} = {\rm odd}$, reproducing the well-known selection rule.
Certainly the effect of the quark substructure goes beyond
the $(1-{\cal P})$ factor in the antisymmetrizer, being also
included through the terms containing $P_{36}$.
\begin{figure}[b]
\vspace*{-0.55cm}
\mbox{\psfig{figure=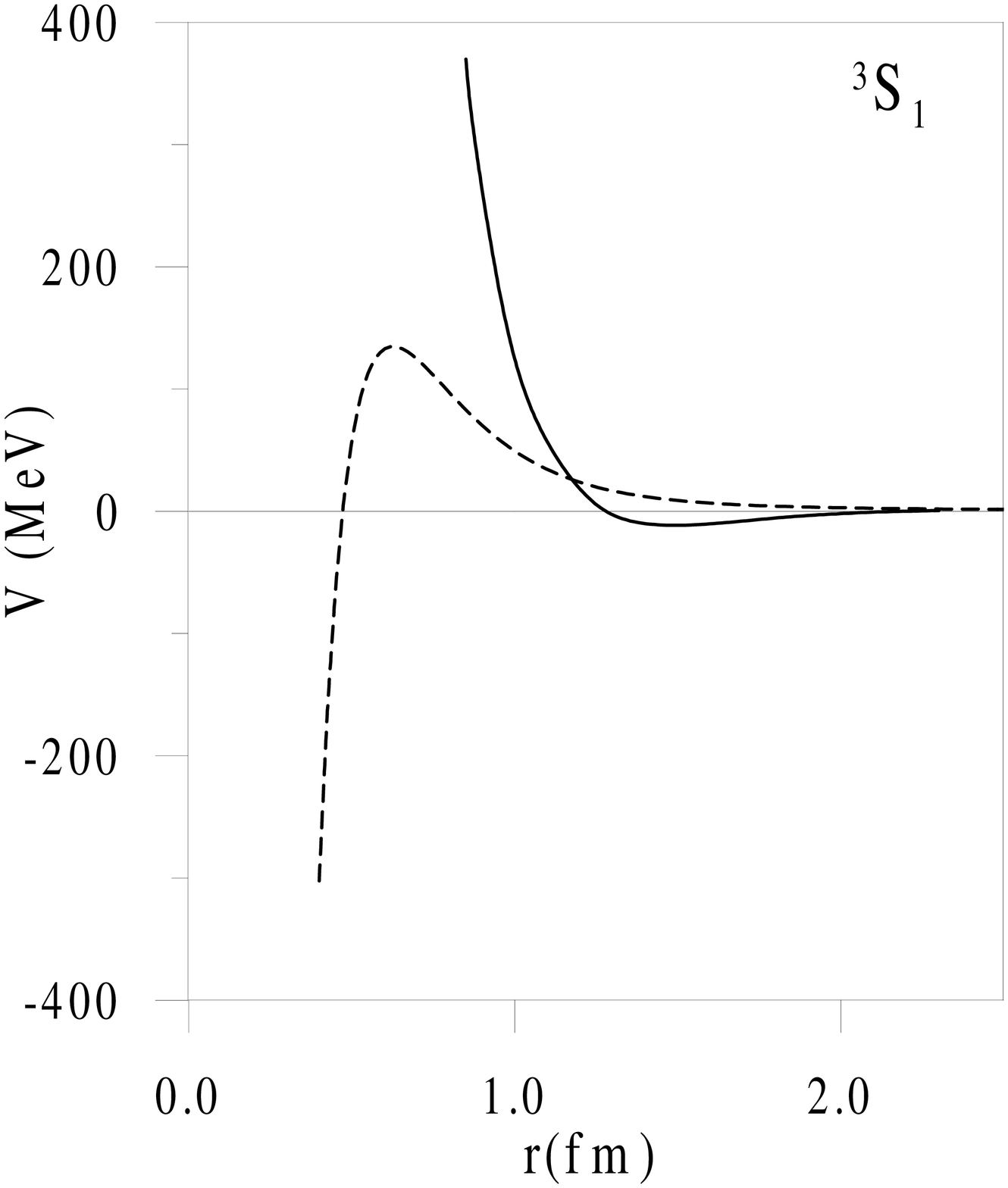,height=2.8in,width=2.2in}}
\hspace*{0.5cm}
\mbox{\psfig{figure=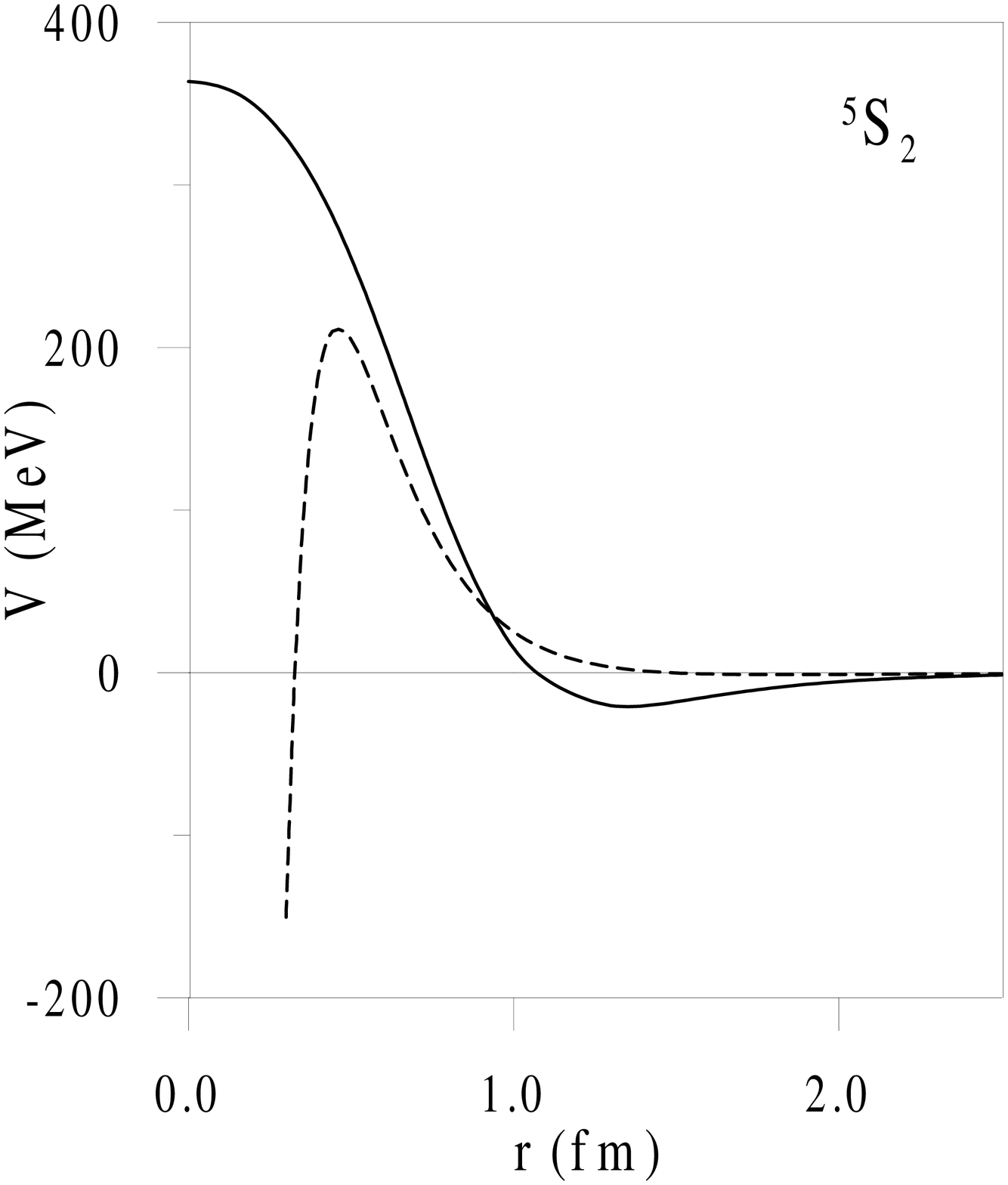,height=2.8in,width=2.2in}}
\vspace*{-1.0cm}
\caption{$^3S_1(T=1)$ and $^5S_2(T=1)$ $N\Delta$ potentials} 
\label{fig2a}
\end{figure}
\begin{figure}[b]
\vspace*{-0.3cm}
\mbox{\psfig{figure=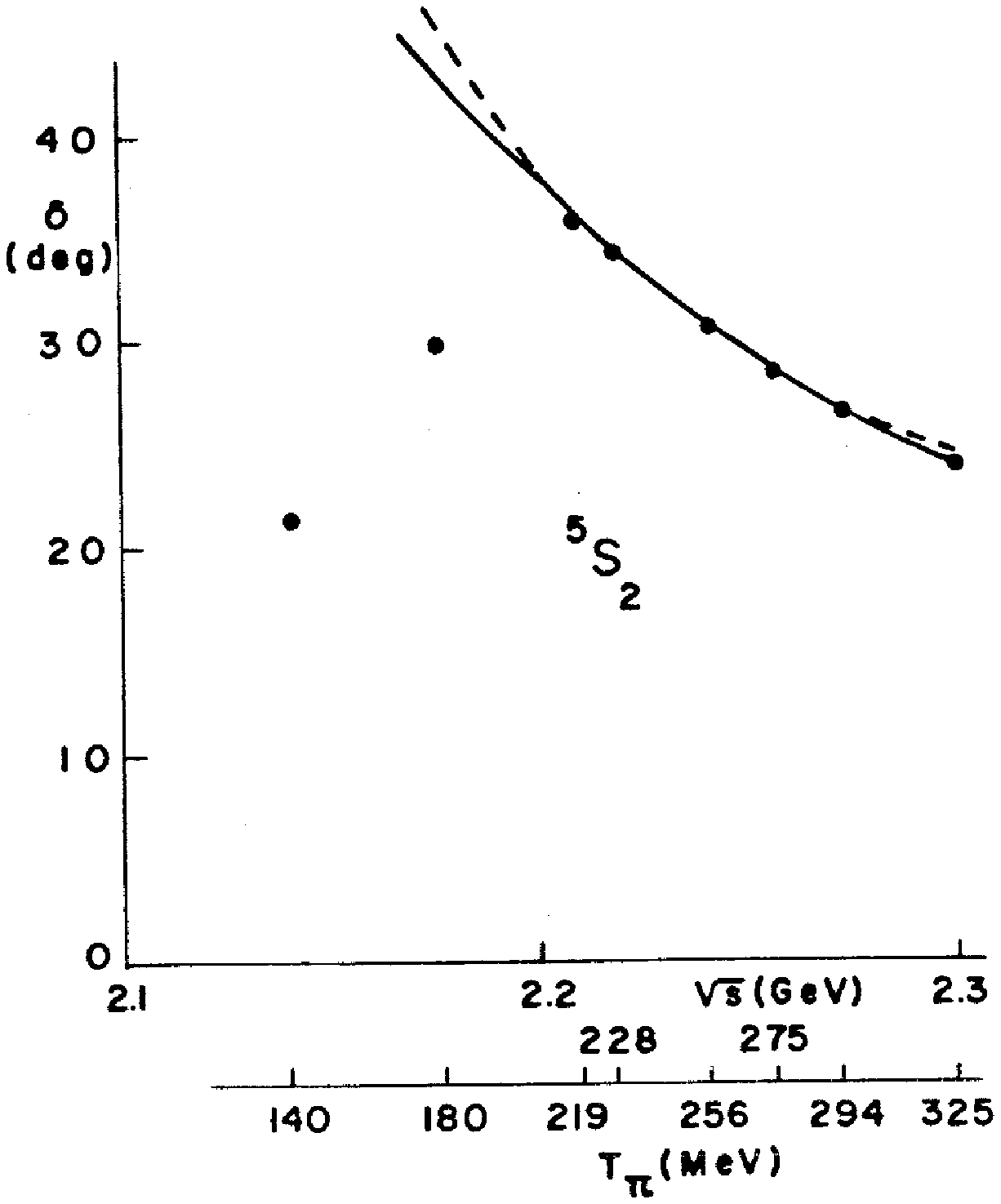,height=2.5in,width=2in}}
\hspace*{0.1cm}
\mbox{\psfig{figure=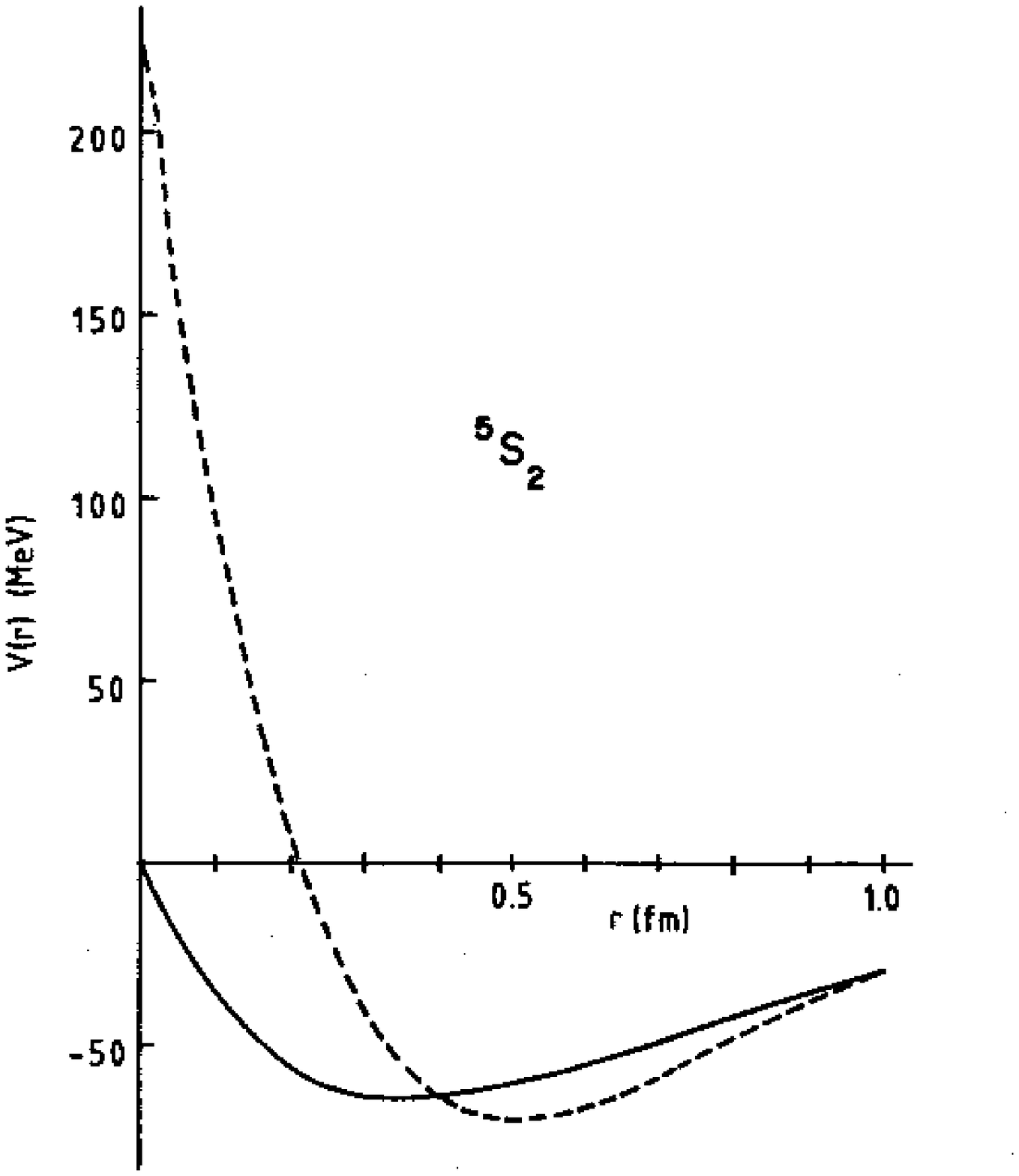,height=2.5in,width=2in}}
\mbox{\psfig{figure=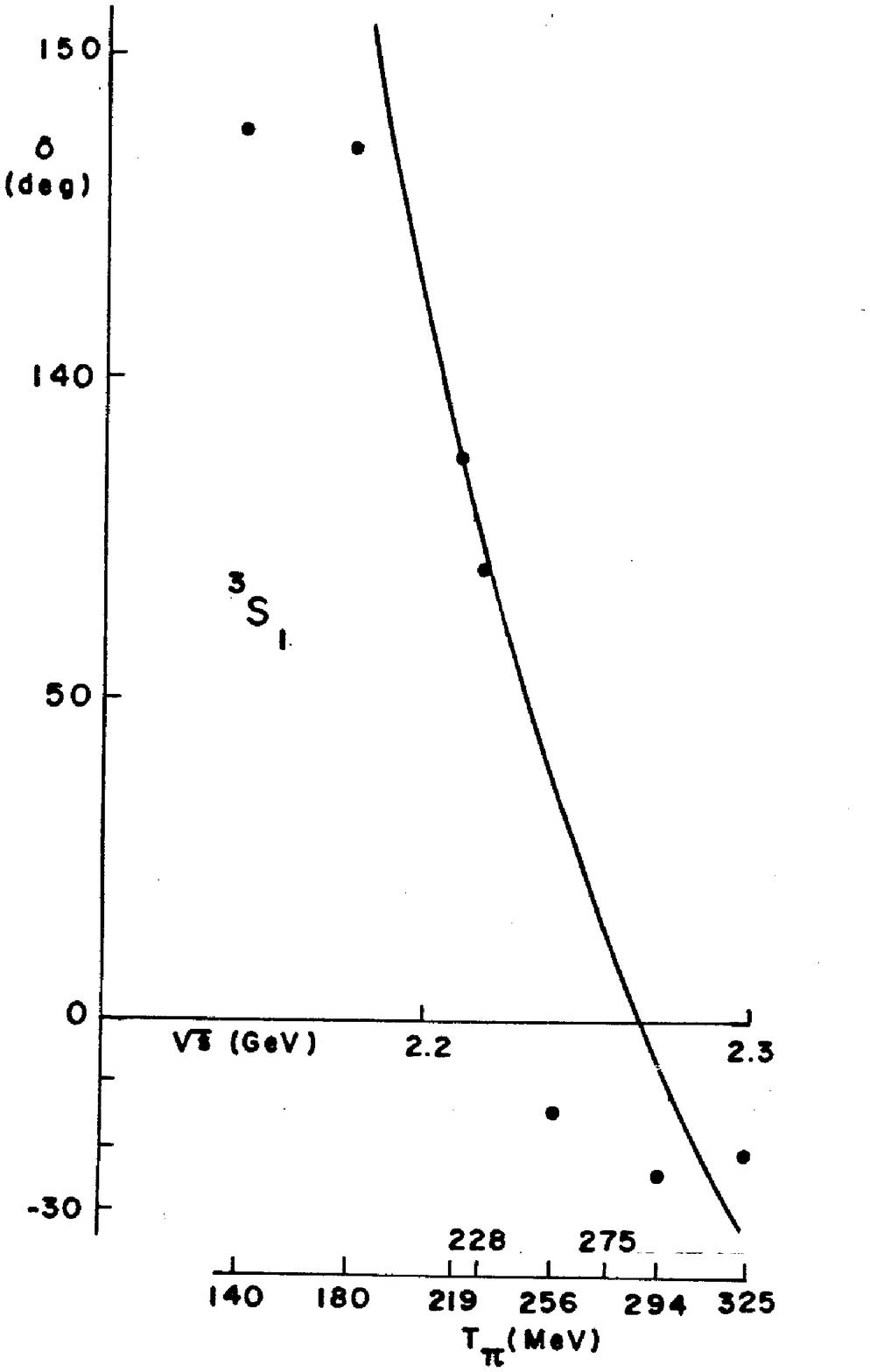,height=2.5in,width=2in}}
\hspace*{0.5cm}
\mbox{\psfig{figure=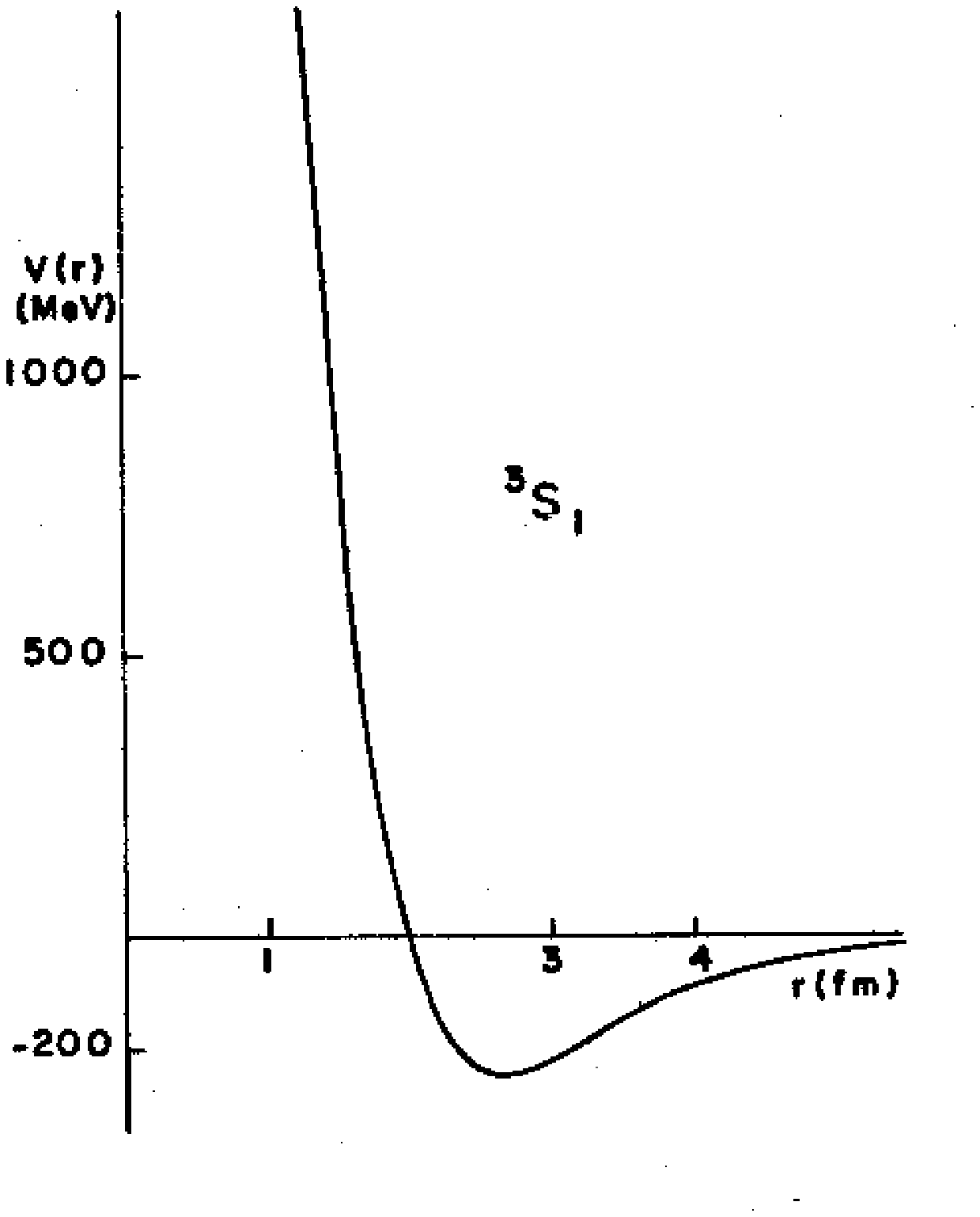,height=2.5in,width=2in}}
\vspace*{-0.5cm}
\caption{$^3S_1(T=1)$ and $^5S_2(T=1)$ $N\Delta$ phase shifts 
and separable potential}
\label{fig2c}
\end{figure}
Let us study the effects of these quark
exchanges in a system of two non-identical baryons,
in particular the $N\Delta$ system. Assuming 
the three-quark cluster wave function 
at position ${\vec S}$ given by 
\begin{equation}
\Phi _B( 123 ;
{\vec S}) \, = \, \prod_{i=1}^{3}
\, \eta_{0s} ({\vec r}_i -{\vec S})  \, \otimes 
\, [ 3 ]_{\rm ST}
\, \otimes \, [ 1 ]_{\rm C} \, ,
\end{equation}
where $\eta_{0s} ({\vec x})$ is
a gaussian with parameter $b$,
and $[ 3 ]_{\rm ST}$ ($[ 1 ]_{\rm C}$) stands for
the spin-isospin (color) representation,
the norm of the projected $\Psi^{{\rm LST}}_{N \Delta}$
state reads,
\begin{equation}
N_{N \Delta}^{{\rm LST}} (R) \, = \, N_L^{\rm di} (R) \,
- \, C(S,T,f) \, N_L^{\rm ex} (R) \, ,
\label{cit1}
\end{equation}
where $N_L^{\rm di}(R)$ [$N_L^{\rm ex}(R)$] refers to the
direct [exchange] radial contribution coming from the
$1 [P_{36}]$ term in the antisymmetrizer. The 
spin-isospin coefficient $C(S,T,f)$ is given by,
\begin{eqnarray}
C(S,T,f) \, & = &  \, 3 \, \left\{
\, _{\rm ST} \langle N (123) \, \Delta (456)
\, \mid \, P_{36}^{ST} \, \mid \,
N (123) \Delta (456) \rangle_{\rm ST} 
\, \, + \, \right. \nonumber \\
& & (-1)^f  \left.
 _{\rm ST}\langle N (123)  \Delta (456)
\mid P_{36}^{ST} \mid 
\Delta (123) N (456) \rangle_{\rm ST} \right\} \, .
\label{cstf}
\end{eqnarray}
\noindent
Its value for
the $N\Delta$ $S$-waves is given in Table \ref{table2} (the $NN$
case is also shown for comparison).
\begin{table}[hbt]
\caption{$C(S,T,f)$ coefficients for the $NN$ and $N\Delta$ $S$-waves.}
\label{table2}
\begin{tabular}{|c|c|c|c|c|c|c|}
\hline
 & \multicolumn{2}{|c|}{$NN$} & \multicolumn{4}{|c|}{$N\Delta$} \\
\hline
 & $^1S_0 $ & $^3S_1 $ & $^3S_1 $ & $^3S_1 $ 
& $^5S_2 $ & $^5S_2 $  \\
 & $(T=1)$ & $(T=0)$ & $(T=1)$ & $(T=2)$ & $(T=1)$ & $(T=2)$ \\ 
\hline
$C(S,T,f)$ & $-$1/9 & $-$1/9 & 1 & 1/9 & 1/9 & 1 \\
\hline
\end{tabular}
\end{table}
In the $R \rightarrow 0$ limit and to the dominant
order Eq. (\ref{cit1}) transforms into:
\begin{eqnarray}
& & N_{N \Delta}^{{\rm LST}} (R) \rightarrow_{ R  \to 0} \, 4 \pi
\left[ {1 - {{3 R^2} \over {4 b^2}}} \right] \,
{ 1 \over {1 \cdot 3 \cdots (2L+1)}} \, \left[ { R^2 \over
{4 b^2}} \right]^L \, \times \nonumber \\
& &  \left\lbrace \left[ 3^L - C(S,T,f) \right] 
+ {1 \over {2(2L+3)}} \left[ { R^2 \over {4 b^2}}
\right]^2  \left[ 3^{L+2} - C(S,T,f) \right]
+ \cdots \right\rbrace
\end{eqnarray}
Then, for $3^L = C(S,T,f)$ the overlapping of the two-cluster
wave function behaves as $R^{2L+4}$
instead of the centrifugal barrier behavior $R^{2L}$, indicating
that quark Pauli blocking occurs (in other words,
a node appears in the relative wave function).
This turns out to be the case
(see Table \ref{table2}) for the $N\Delta$ partial waves $^3S_1(T=1)$
and $^5S_2(T=2)$ giving rise to forbidden states 
(the \se orbital configuration for L=0). 
If we now calculate the $N\Delta$ potential, we observe 
(solid line) the important difference between the potential
in a Pauli blocked
and a non Pauli blocked channel (see Fig. \ref{fig2a}). 
At the same time we have represented
by the dashed line a $N\Delta$ potential obtained from a direct
scaling of the $NN$ interaction \cite{PEHE} and we see how the 
difference between the two channels is just a spin-isospin
factor. As shown in Fig. \ref{fig2c}, 
a separable $N\Delta$ potential reproducing
the phase shifts obtained from the
experimental $\pi d$ elastic cross section presents 
the strong repulsion predicted by the
quark model in the $^3S_1(T=1)$ partial wave \cite{FEDO}.
 
A similar analysis for the $NN$ system does not show
any forbidden configuration, see Table \ref{table2}.
As we have seen, the $NN$ short-range repulsion in the S-waves
involves dynamical effects.

\subsection{Connection to Other Systems: The Baryon Spectrum}
The chiral constituent quark model allows also
to study the baryon spectrum
in a completely parameter-free way.
We have calculated the nonstrange baryon spectrum
within a Faddeev approach. We included all
the configurations $(\ell,\lambda,s,t)$ 
($\ell$ is the orbital angular momentum of a pair, $\lambda$ 
is the orbital angular momentum between the pair and the third 
particle, while $s$ and $t$ are the spin and isospin of the pair) 
with $\ell$ and $\lambda$ up to 5. 
 
\begin{figure}[t]
\vspace*{-1.1cm}
\mbox{\psfig{figure=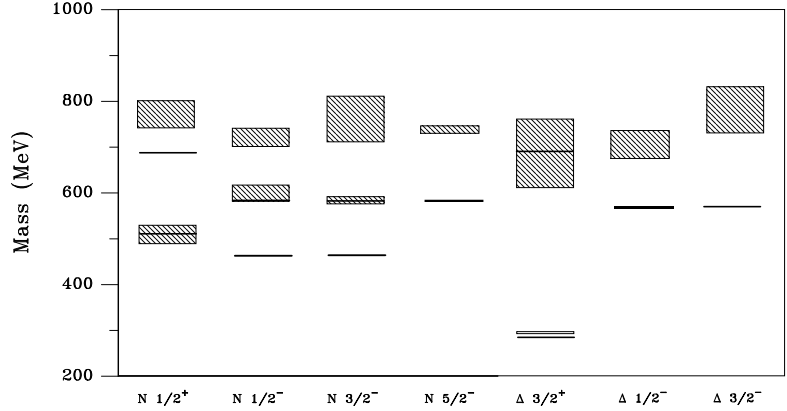,height=3.3in,width=4.9in}}
\vspace*{-2.6cm}
\caption{Relative energy nucleon and $\Delta$ spectra} 
\label{fig5a}
\end{figure}

For this purpose, the 
delta function of the OGE interaction has to be regularized. 
By choosing an exponential regularization
\begin{equation}
\delta({\vec r}_{ij}) \, \Rightarrow \,
{1 \over {4 \pi r_0^2}} \,\,
{{e^{-r_{ij}/r_0}} \over r_{ij}} \, ,
\end{equation}
with $r_0 = 0.25$ fm (a typical value
for spectroscopic models) 
we obtain the $N$ and $\Delta$ spectra shown in Fig. \ref{fig5a}.
The predicted spectrum is reasonable except for the
fact that all the negative parity states are around 100 MeV
below the experimental data. 
The relative position of the $N(1440)$ and the $N(1535)$ is not very
much affected by the modification of $\alpha_s$ (see
Fig. 3 of ref. \cite{GAR1}),
indicating that the correct level ordering 
is not connected with the strength of OGE.

\begin{figure}[b]
\vspace*{-0.5cm}
\centerline{\psfig{figure=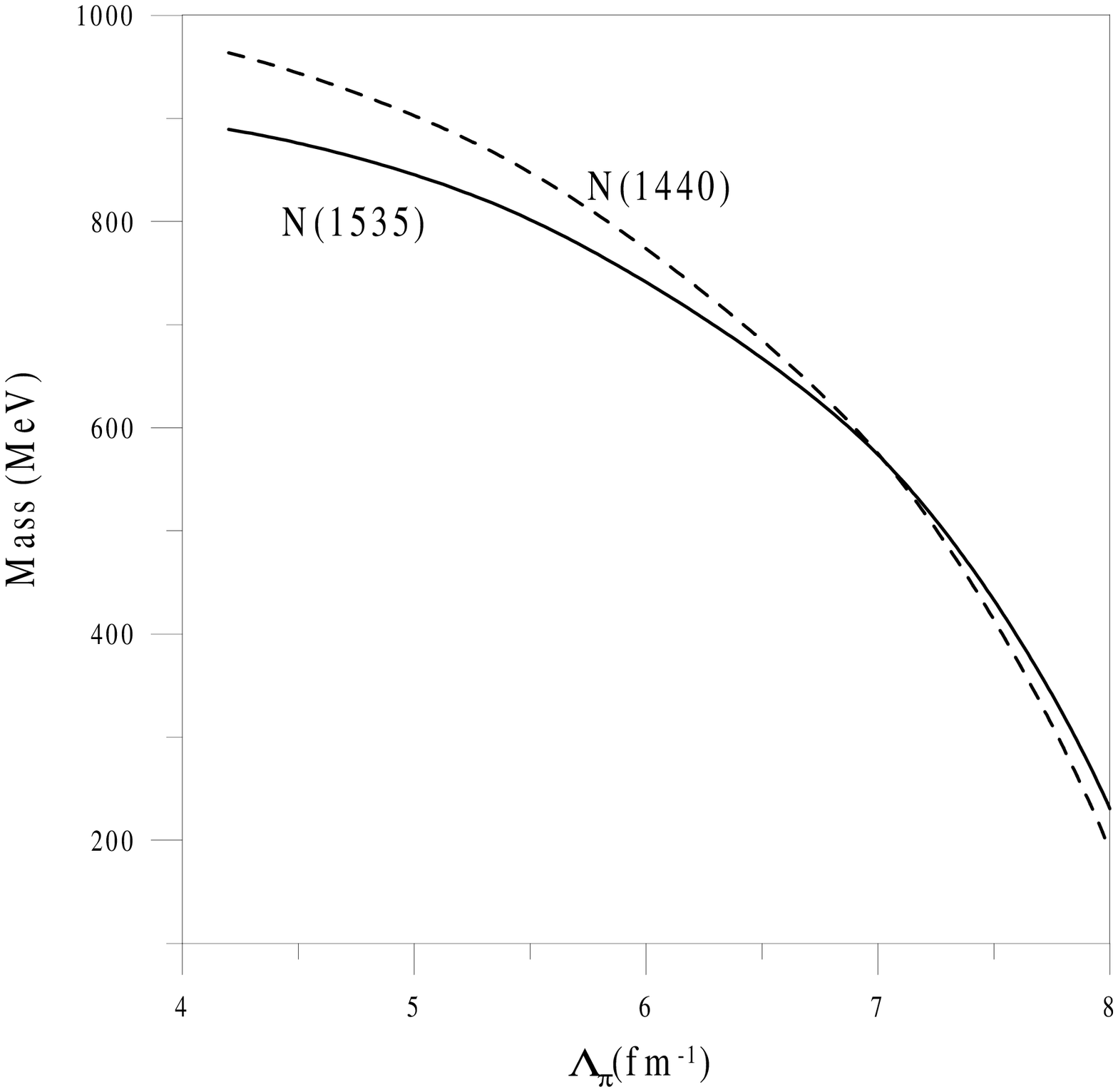,height=2.8in,width=2.2in}}
\vspace*{-2.7cm}
\caption{$N(1440)$ and $N(1535)$ masses as a function of $\Lambda_\pi$}
\label{fig5b}
\end{figure}

A possible mechanism
responsible for the inversion of these states can be obtained through
the pseudoscalar interaction. To illustrate this point we have
calculated the energy of the $N(1440)$ and $N(1535)$ increasing
the contribution of the pseudoscalar interaction by
letting the cutoff parameter $\Lambda_\pi$ of the OPE to increase.
The results are shown in Fig. \ref{fig5b}. As can be seen, the inversion of
the ordering between the positive and negative parity states
can be achieved if $\Lambda_\pi$ becomes sufficiently large
(around 7 fm$^{-1}$ for the set of parameters of Table \ref{table2}). 
Such a model would be incompatible with the understanding of the
basic features of the two-nucleon system \cite{TOK2}.
\section{Summary}

In this work we have emphasized several physical aspects 
of the $NN$ interaction that
support the results obtained within the chiral constituent quark
model. In particular, medium range attraction is satisfactorily
explained and the $NN$ short-range repulsion is understood
through antisymmetry effects on the OPE interaction. Actually,
antisymmetry effects at the level of Goldstone bosons appear as a basic
tool to understand the $NN$ phase shifts. Extension of the model
to the general baryon-baryon case is straightforward and parameter-free.
From it, non-strange meson-baryon-baryon coupling constants can be
calculated.
Pauli blocking effects manifest in different partial waves of several
systems and find support in indirect experimental data. Such strong
repulsion cannot be understood in a meson-exchange model.

Finally, the chiral constituent quark model 
is able to generate a quite reasonable description of the baryon
spectrum with a set of parameters that allows to understand the
$NN$ phenomenology. 
Therefore, the validity of a model where
the OGE is combined with Goldstone-boson exchanges is out of
question, and it presents advantages with respect to models
based only on Goldstone-boson exchanges,
as has been
recently emphasized in several works \cite{TOK2,ISGU}.

\begin{acknowledge}
This work has been partially funded by
Ministerio de Ciencia y Tecnolog{\'\i}a
under Contract No. BFM2001-3563, by Junta de
Castilla y Le\'{o}n under Contract No. SA-109/01,
and by the European Comission IHP program
under Contract No. HPRN-CT-2000-00130.
\end{acknowledge}

\end{document}